\documentclass[sigconf, preprint]{acmart}

\AtBeginDocument{%
  \providecommand\BibTeX{{%
    \normalfont B\kern-0.5em{\scshape i\kern-0.25em b}\kern-0.8em\TeX}}}

\setcopyright{acmcopyright}
\copyrightyear{2023}
\acmYear{2023}

\acmConference[Conference acronym 'XX]{Make sure to enter the correct
  conference title from your rights confirmation email}{XXXX xx--xx,
  20xx}{XXX, XXX}
\definecolor{brickred}{rgb}{1,0,0}
\newcommand{\red}[1]{{\color{brickred}{#1}}}

\usepackage[ruled,linesnumbered]{algorithm2e}
\usepackage{tabularx}
\usepackage{algorithmic}

\begin{document}

\title{Bridging the Information Gap Between Domain-Specific Model and General LLM for Personalized Recommendation}

\author{Wenxuan Zhang}
\email{zwx980624@gmail.com}
\affiliation{%
  \institution{Academy for Advanced Interdisciplinary Studies, Peking University}
  \city{Beijing}
  \country{China}
}

\author{Hongzhi Liu}
\email{liuhz@pku.edu.cn}
\affiliation{%
  \institution{School of Software and Microelectronics, Peking University}
  \city{Beijing}
  \country{China}}

\author{Yingpeng Du}
\email{dyp1993@pku.edu.cn}
\affiliation{%
  \institution{School of Software and Microelectronics, Peking University}
  \city{Beijing}
  \country{China}
}

\author{Chen Zhu}
\email{zc3930155@gmail.com}
\affiliation{%
 \institution{Career Science Lab, BOSS Zhipin}
 \city{Beijing}
 \country{China}}

\author{Yang Song}
\email{songyang@kanzhun.com}
\affiliation{%
  \institution{NLP Center, BOSS Zhipin}
  \city{Beijing}
  \country{China}}

\author{Hengshu Zhu}
\email{zhuhengshu@gmail.com}
\affiliation{%
  \institution{Career Science Lab, BOSS Zhipin}
    \city{Beijing}
  \country{China}}

\author{Zhonghai Wu}
\email{wuzh@pku.edu.cn}
\affiliation{%
  \institution{School of Software and Microelectronics, Peking University}
  \city{Beijing}
  \country{China}}

\renewcommand{\shortauthors}{Wenxuan Zhang, et al.}

\begin{abstract}


Generative large language models~(LLMs) are proficient in solving general problems but often struggle to handle domain-specific tasks. This is because most of domain-specific tasks, such as personalized recommendation, rely on task-related information for optimal performance. 
Current methods attempt to supplement task-related information to LLMs by designing appropriate prompts or employing supervised fine-tuning techniques. 
Nevertheless, these methods encounter the certain issue that information such as community behavior pattern in RS domain is challenging to express in natural language, which limits the capability of LLMs to surpass state-of-the-art domain-specific models. On the other hand, domain-specific models for personalized recommendation which mainly rely on user interactions are susceptible to data sparsity due to their limited common knowledge capabilities.
To address these issues, we proposes a method to bridge the information gap between the domain-specific models and the general large language models. Specifically, we propose an information sharing module which serves as an information storage mechanism and also acts as a bridge for collaborative training between the LLMs and domain-specific models. By doing so, we can improve the performance of LLM-based recommendation with the help of user behavior pattern information mined by domain-specific models. On the other hand, the recommendation performance of domain-specific models can also be improved with the help of common knowledge learned by LLMs. Experimental results on three real-world datasets have demonstrated the effectiveness of the proposed method.

\end{abstract}







\maketitle

\section{Introduction}
With the explosion of information, humans have become increasingly reliant on recommendation systems~(RS) across various domains, such as e-commerce, social media, personalized healthcare, etc \cite{bobadilla2013recommender}. Recently, the rapid development of generative large language models~(LLMs) \cite{zhao2023survey} has provided ample world-level knowledge and reasoning capabilities, thereby opening up new paradigms for recommendation systems \cite{wu2023survey}.

\begin{figure}[t]
    \centering
      \includegraphics[width=8.5cm]{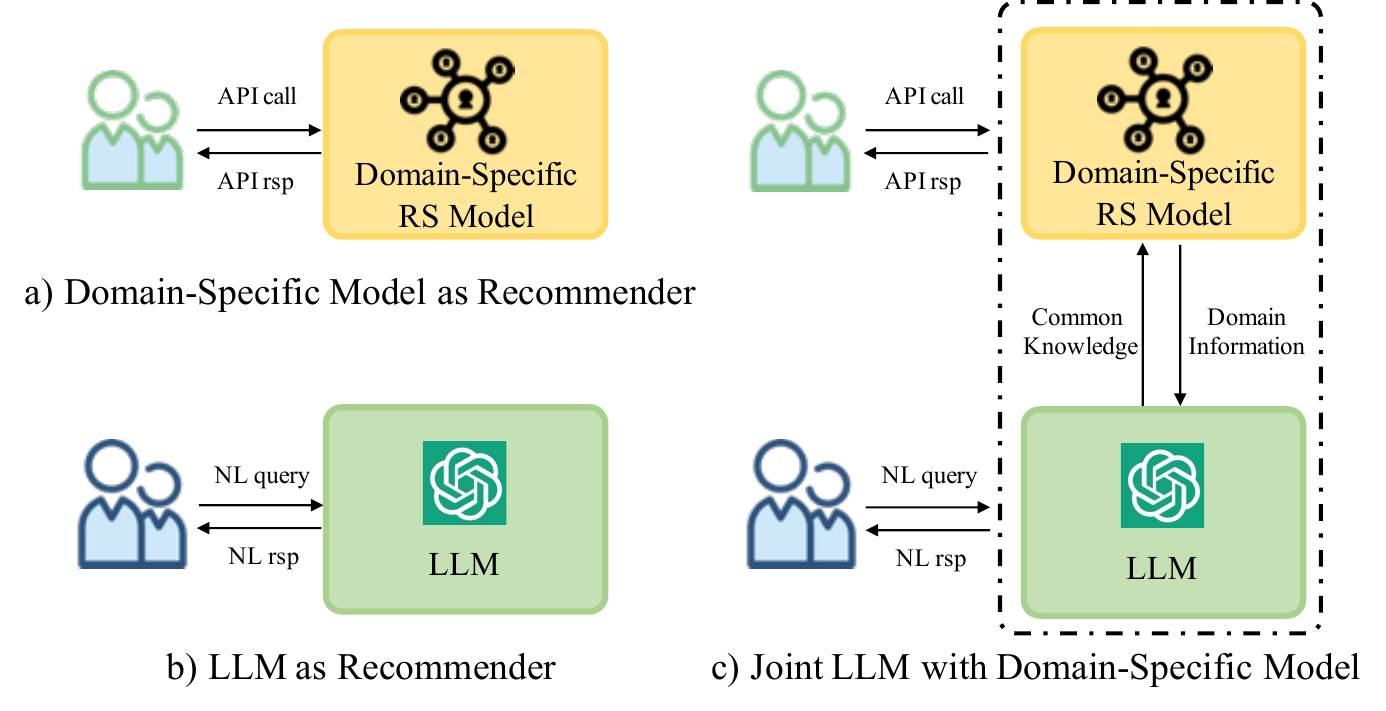}
    \caption{Paradigms of recommendation systems. a) a domain-specific model serves as the recommender. b) LLM takes on the role of the recommender. c) our proposed paradigm, a unified framework combining both LLM and domain-specific model.}
    \label{fig:first}
\end{figure}


Traditional recommenders based on domain-specific models which try to learn users' preferences based on historical interactions and generates personalized item lists tailored to individual users, as shown in Figure~\ref{fig:first} a). With the advancement of LLMs, another recommendation paradigm emerges, which takes the LLM as the recommendation system itself, as shown in Figure 1 b). By processing user dialogues or instructions expressed in natural language, the LLM can also generate personalized recommendations with the help of a wealth of general knowledge and reasoning abilities pre-trained from vast amounts of data. Subsequently, these recommendations are conveyed to users in a natural language format.

\begin{figure}[t]
    \centering
      \includegraphics[width=8.5cm]{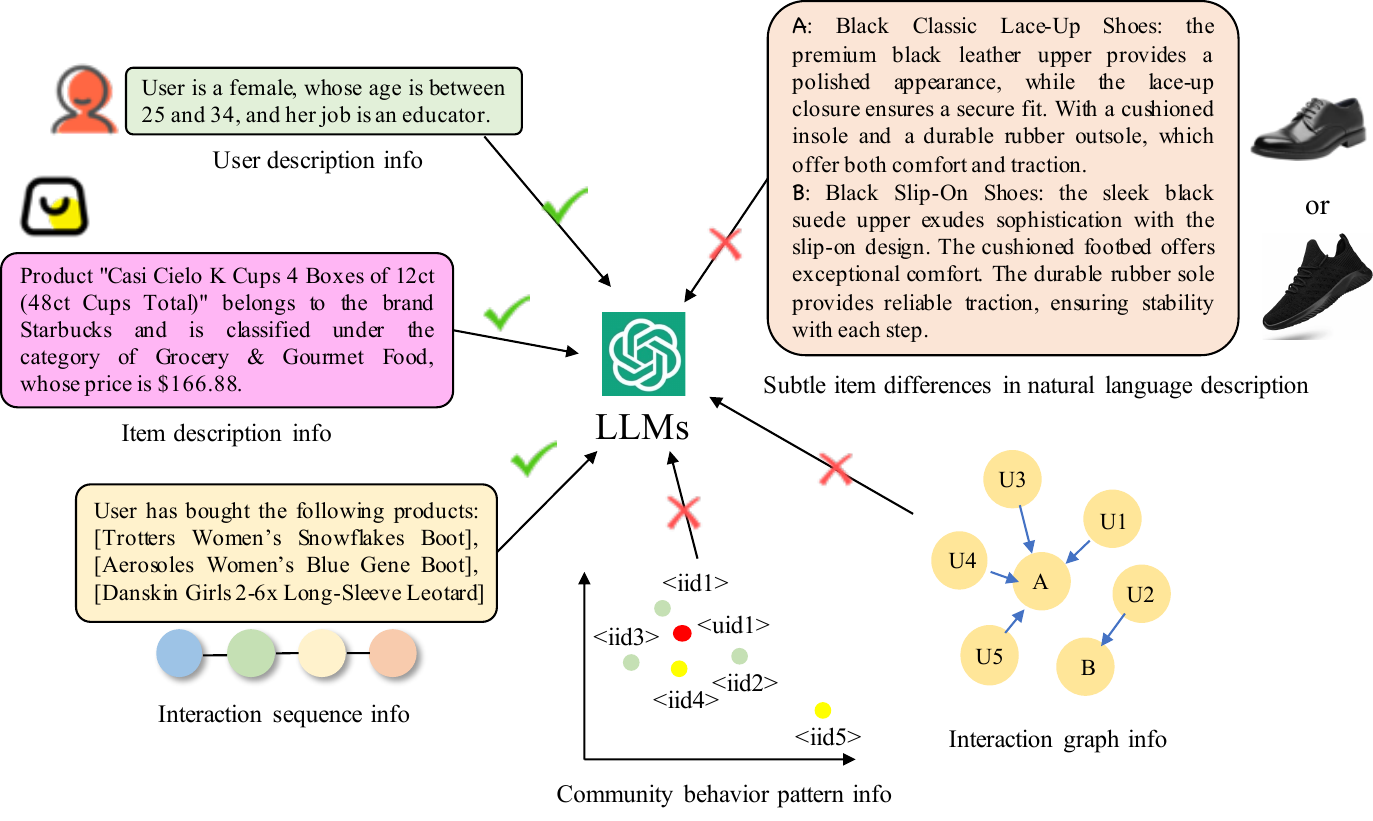}
    \caption{Illustration for text-is-not-all-your-need, where user description, item description and interaction sequences can be expressed in text form and convey to LLM while subtle item differences, community behavior patterns and interaction graph are hard to convey to LLM.}
    \label{fig:prompts}
\end{figure}

However, both these two paradigms has their own challenges. General-purpose LLMs often lack domain-specific information which results in poor performance compared to state-of-the-art domain-specific models. Firstly, some domain-specific information for RS is hard to convey to LLM by prompt engineering. In recommendation systems based on LLMs, all information about items and users can only be conveyed to the LLM through natural language. As shown in Figure~\ref{fig:prompts}, existing methods can express various recommendation domain information in textual format \cite{li2023text}. However, this "text-is-all-your-need" way may not always work well for all kinds of domain-specific information. To be specific:

\begin{itemize}
    \item [1)] It is challenging for natural language to distinguish among similar products with subtle differences.
    \item [2)]The community behavior pattern information in the field of recommendation systems is difficult to express in natural language.
\end{itemize}

Moreover, domain-specific information for RS is hard to learn for LLM by supervised fine-tuning~(SFT). LLMs may be good at learning frequent patterns which could be formalized as knowledge, but it is difficult for them to mine and make use of community behavior patterns such as local $k$-nearest neighborhood information. Overall, conveying such information to LLMs becomes challenging. However, domain-specific models, for instance NCF and lightGCN, can utilize flexible model structures, data structures, and objective functions to model and utilize this kind of information. 

Furthermore, the paradigm of traditional domain-specific model as recommender also has its flaws. In many recommendation scenarios, the interaction data between users and items is often sparse. Most domain-specific models that mainly rely on user interactions cannot perform well in such scenarios. However, general large language models possess robust common knowledge and reasoning capabilities, allowing them to address the issue of sparse data by extracting personalized intentions and preference information from user contexts and side information from item descriptions.

To take advantages of both these two paradigms, we propose a joint framework to \textbf{B}ridge the information gap between \textbf{D}omain-specific models and generative \textbf{L}arge language \textbf{M}odels~(BDLM) for personalised recommendation. The proposed new paradigm of RS is illustrated in the Figure~\ref{fig:first} c). To be Specific, we design an information sharing module to transfer information between domain-specific models and LLMs for RS task. Furthermore, we design a deep mutual learning strategy to jointly training the LLM and the domain-specific model to enhance the information sharing loop and ultimately improve training effectiveness.

The main contributions of this paper can be summarized as follows. 

1) To the best of our knowledge, this is the first work to propose a unified paradigm for bridging the information gap between domain-specific models and generative large language models for personalized recommendation. Therefore, we can improve the recommendation performance of both the generative LLMs and the domain-specific models.

2) Joint training, employing the strategy of deep mutual learning between LLMs and domain-specific models, enhances the training effectiveness . A cascaded structure comprising an LLM and a domain-specific model can be utilized for inference, enabling both the LLM and the domain-specific model to achieve state-of-the-art performance.

3) Eperiments on multiple recommendation datasets from various domains validate the effectiveness of the proposed approach.
    

\section{Related Work}


\subsection{Recommendation System}
In order to address the issue of information overload, recommendation systems have emerged as essential tools in delivering personalized content to users across diverse domains \cite{bobadilla2013recommender}. Common recommendation methods can be divided into two main categories: interaction-based recommendation and content-based recommendation \cite{thorat2015survey}. The core idea behind interaction-based recommendation algorithms is to uncover users' behavior patterns from their historical interactions. One of the most classic algorithms in this category is collaborative filtering~(CF) \cite{su2009survey, linden2003amazon, sarwar2001item}, which assumes that users with similar behaviors have similar preferences.

With the development of neural recommendation systems \cite{zhang2019deep}, embedding representations have become an important method for modeling user preferences and item characteristics. The Matrix Factorization (MF) \cite{koren2009matrix} method decomposes the user-item interaction matrix into user and item embedding representations, forming the prototype of embedding-based methods. Subsequently, the Neural Collaborative Filtering (NCF) \cite{he2017neural} method utilizes the embedding vectors of users and items to implement collaborative filtering. The trained embedding vectors represent the users' behavior patterns and the characteristics of items in a high-dimensional space. Furthermore, representation learning methods based on user-item interaction graphs emerge, which can better capture high-order neighbor information of users and items. For example, lightGCN \cite{he2020lightgcn} utilizes graph convolutional network~(GCN) \cite{kipf2016semi} to extract high-order neighbor information from the user-item interaction graph, which contains more comprehensive user behavior pattern information.

Content-based methods, on the other hand,  utilize supplementary contextual information to model the level of compatibility between users and items. This contextual information includes user-specific information such as personal profiles, interests, as well as item attributes like categories and descriptions. In the scenarios where user and item interaction data is sparse, context-based methods often exhibit better recommendation performance. ReBKC \cite{hui2022personalized} extracts auxiliary information like social networks and item properties from knowledge graphs to enhance recommendation performance. U-BERT \cite{qiu2021u} employs a BERT-like model customized for recommendation tasks. By leveraging pre-training and fine-tuning techniques, U-BERT enhances its ability to comprehend and utilize user contextual information, thereby effectively capturing and modeling user preferences.

\subsection{LLMs for Recommendation}

As an emerging NLP technology, LLMs possess billions of parameters and undergo extensive pre-training \cite{raffel2020exploring, zhang2022opt, ouyang2022training}. With a wealth of common knowledge and reasoning abilities, LLMs showcase remarkable performance across various tasks. In the field of personalized recommendation, the introduction of LLMs serves two main purposes. Firstly, LLMs leverage the emerging capabilities to serve as a content-based recommender and directly tackle recommendation tasks. Secondly, LLMs harness their knowledge and reasoning capabilities to furnish information to domain-specific models, thereby amplifying their recommendation performance.

LLM-Rec \cite{lyu2023llm} and LLM-CR \cite{sanner2023large} directly harness the emerging capabilities of LLMs by designing appropriate prompts. This enables the LLM to undertake tasks such as top-K recommendations, sequence recommendations, and providing explanations for recommendations. To further enhance the performance of LLMs in recommendation tasks, InstructRec \cite{zhang2023recommendation} employs SFT methods to fine-tune the LLMs, thereby further improving their recommendation abilities. LGIR \cite{du2023enhancing} utilizes the summarization and reasoning capabilities of LLMs. It leverages the user's resume and the descriptions of historically interacted jobs to generate new user resumes, which has shown improvements in personalized job recommendation tasks.

However, all these LLM-based methods fail to effectively extract community behavior pattern information such as $k$-nearest neighborhood information from interaction graph. These information can be extracted by the domain-specific model from a large amount of user-item interaction behavior. Thus, incorporating this information into LLM will greatly enhance the recommendation performance of LLM-based methods.

\section{Problem Definition}

Let $\mathcal{U}=\{ u_1, ..., u_N \}$ and $\mathcal{I} = \{ i_1, ..., i_M \}$ represent the sets of $N$ users and $M$ items, respectively. The interaction records between users and items can be denoted as an interaction matrix $\mathcal{R} \in \mathbb{R}^{N \times M}$, where

\begin{equation}
\begin{split}
    \mathcal{R}_{jk} = \left \{
\begin{array}{lr}
    1,                    & \mathrm{user\ } u_j \mathrm{\ interated\ with\ item\ } i_k\\
    0,                                 & \mathrm{otherwise}
\end{array}
\right.
\end{split}
\end{equation}

Each interaction $\mathcal{R}_{ui}$ can be converted to a prompt or instruction $C_{ui} = [w_1,...w_{l_c}]$ based on user's contexts and item's side information, where $w_i$ is the $i$-th word in the prompt $C_{ui}$ and $l_c$ denotes the length of the prompt $C_{ui}$. 

The goal of RS task is to recommend proper items to target users. Formally, we need to construct a matching function $g(u, i, C_{ui})$ based on the interaction records $\mathcal{R}$ and the prompt $C_{ui}$, and then make the top-K recommendation or interaction prediction based on the matching function.

\section{The Proposed Method}

\begin{figure}[t]
    \centering
    \includegraphics[width=8.5cm]{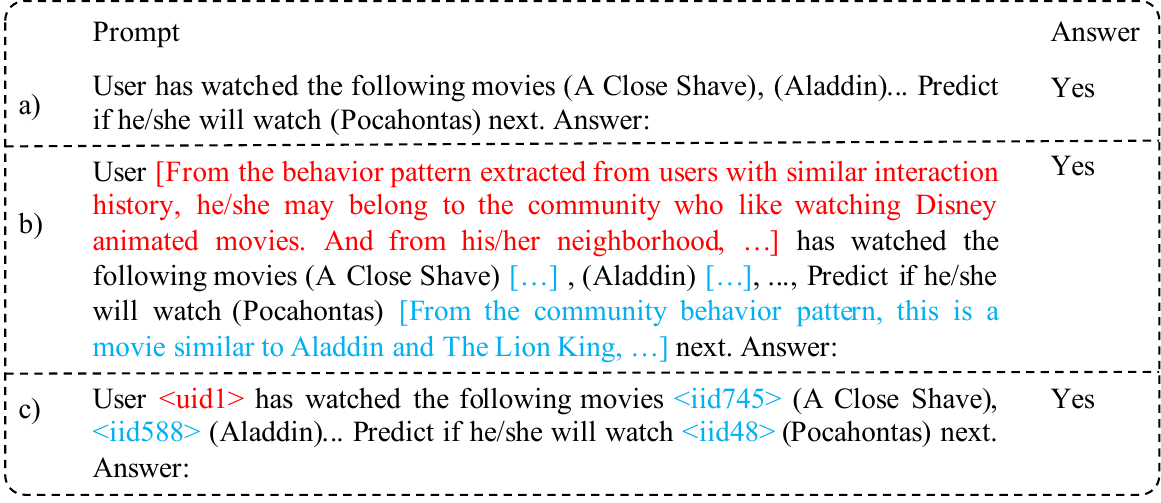}
    \caption{A case demonstrating the motivation of designing task-specific tokens.}
    \label{fig:prompt}
\end{figure}

\begin{figure*}[t]
    \centering
    \includegraphics[width=17.5cm]{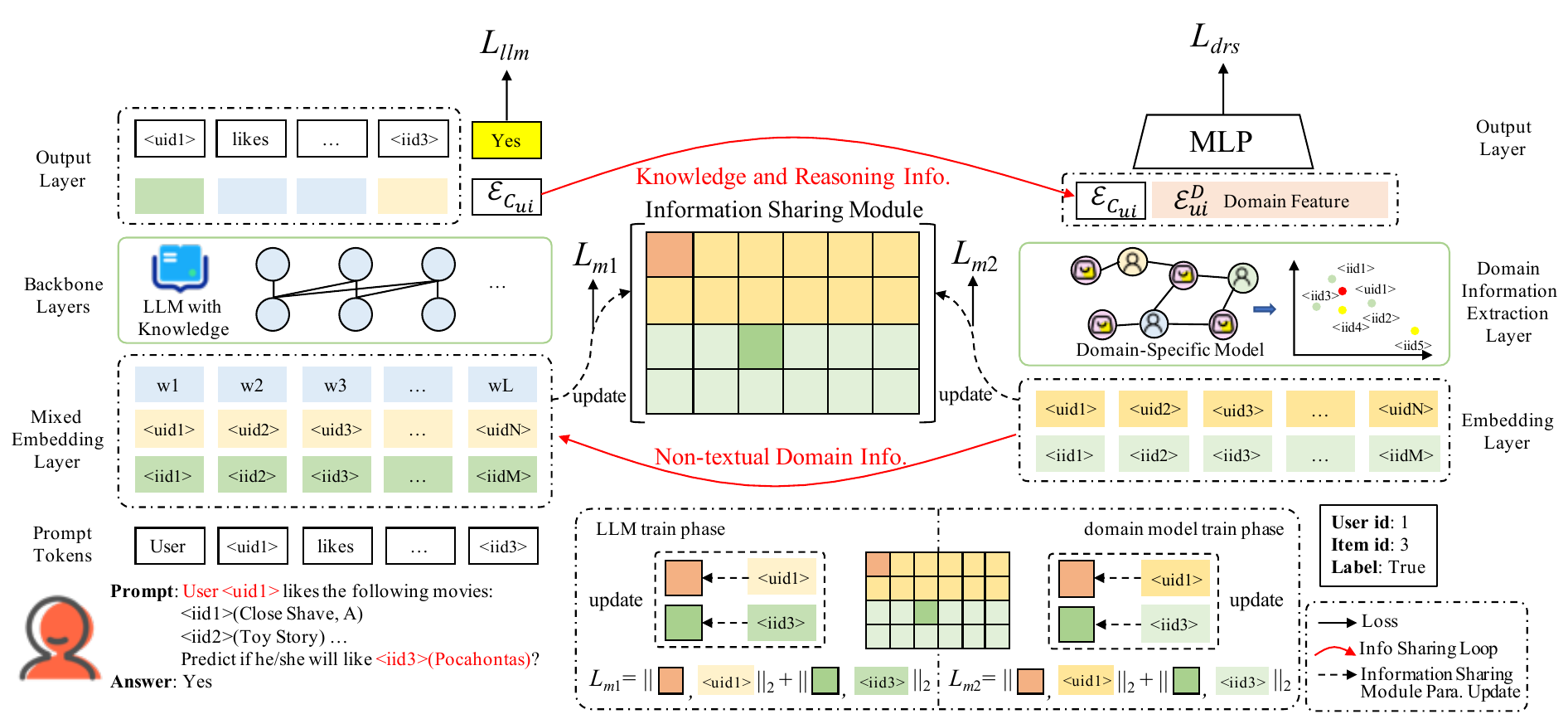}
    \caption{An illustration of the proposed architecture}
    \label{fig:main}
\end{figure*}

To address the limitations of LLMs and domain-specific models while taking their advantages, we design a unified framework that bridge the information gap between a LLM and a domain-specific recommendation model. Its overall architecture is illustrated in Figure~\ref{fig:main}. 

First, we design a domain information enhanced LLM-based recommendation module to solve the domain information deficiency problem of LLM. Then, we propose a common knowledge enhanced recommendation module for solving the sparse problem in domain-specific model. Finally, a joint learning between LLM and domain-specific model based on deep mutual learning is proposed for better enhancing the information sharing loop.

\subsection{Domain Information Enhanced LLM-Based Recommender}

\subsubsection{Design of Task-Specific Tokens}

The general LLMs like ChatGPT\footnote{https://openai.com/blog/chatgpt} and LLaMA \cite{touvron2023llama} are trained on large data of multiple domains which are designed to address general problems, but it may not perform well when applied to domain-specific tasks such as personalized recommendation. To improve the performance of LLM in domain-specific tasks, a common approach is to perform task-specific supervised fine-tuning~(SFT). The core of SFT is the design of training samples, each of which consists of a prompt $C_{ui}$ and an answer $A_{ui}$.

A common prompt for personalized recommendation can be structured as follows: 
$\mathtt{prompt}(\mathrm{user\_description}, \mathrm{item\_description}, \\ \mathrm{task\_form})$, 
where all user descriptions, item descriptions and task forms are expressed in natural language to provide domain information. As the case shown in Figure~\ref{fig:prompt} a), the prompt can provide  basic information like user's historical behavior and item's name.


Nevertheless, this basic prompt exhibits several limitations. Firstly, it lacks the ability to differentiate between users, resembling more of a session-based recommendation approach. Additionally, it poses challenges in distinguishing between items that have similar descriptions. If we attempt to use identifiers such as "uid\_1" or "iid\_2" to differentiate users and items, they will be tokenized into multiple sub-word units like ["uid", "\_", "1"] using natural language vocabulary $\mathcal{V}$, thereby losing their discriminative capability to some extent. Moreover, only basic information provided may not be adequate for an LLM-based recommender to generate accurate predictions. As shown in Figure~\ref{fig:prompt} b), by incorporating more effective domain information for personalized recommendation into the prompt, LLM can attain a significantly higher level of performance. However, summarizing this kind of information into natural language is challenging using existing methods. 

Therefore, we adopt an approach of introducing external task-specific special tokens of RS, such as <uid1> and <iid2>, as depicted in Figure~\ref{fig:prompt} c). By utilizing these tokens, LLM becomes more adept at distinguishing users and items and can effectively integrate non-textual domain information through pre-loading and fine-tuning their embeddings.

To be specific, let $\mathcal{V}$ represent tokens for natural language. $\mathcal{U}_{tok}$ represents user tokens from "<uid1>" to "<uid$N$>" and $\mathcal{I}_{tok}$ represents item tokens from "<iid1>" to "<iid$N$>". Thus, the new vocabulary $\mathcal{W} = \mathcal{V} \cup \mathcal{U}_{tok} \cup \mathcal{I}_{tok}$. The prompt tokens $w_j$ in the new prompt $C_{ui}$ belong to $\mathcal{W}$.


\subsubsection{Domain Information Enhanced LLM}





Based on the idea to incorporate domain information by task-specific tokens. We propose an embedding-based domain information enhanced LLM, which consists of a mixed embedding layer and trasformer-based backbone layers.

We propose a mixed embedding layer containing special user/item tokens, which enable us to integrate user and item embeddings into the LLM computation, denoted as $U_e^L$ and $I_e^L$.

Formally, the mixed embedding layer accept an input prompt $C_{ui} = [w_1, ..., w_{l_c}]$. We obtain the embedding vector $E_{C_{ui}}$ by:

\begin{equation}
\begin{split}
    &W_e = \mathtt{concat}(V_e; U_e^L; I_e^L) \\
    &E_{C_{ui}} = \mathtt{one\_hot}({C_{ui}}) W_e.
\end{split}
\end{equation}

However, the training of the mixed embedding layer presents the following challenges:

Firstly, the introduction of newly added tokens without pre-training are hard to convergence and adaptability to the pre-existing trained LLM. Secondly, the magnitude of users and items in real-world RS can be significantly large. Consequently, fine-tuning these tokens from scratch would require a substantial amount of training data and resources investment.

Therefore, we propose to initialize $U_e^L$ and $I_e^L$ with the corresponding embedding $DRS_u^0$ and $DRS_i^0$ from a domain-specific model, such as NCF and lightGCN. By doing so, we can not only accelerate the convergence of these newly added tokens, leverage training data, but transmit non-textual domain information as well.


Following the mainstream LLM structure \cite{radford2018improving}, we employ an n-layer transformer decoder as the backbone of our large language model. Each transformer layer consists of a multi-headed self-attention followed by a layer normalization and a MLP. The backbone layers receive token embeddings of the prompt $E_{C_{ui}}$ as input and generate an encoded representation layer by layer. The output of the final backbone layer is fed into the output layer, which is a position-wise feed-forward network to generate an output distribution over target tokens.



\subsubsection{SFT Training Procedure}
We incorporate the newly added user and item tokens to form training instances as shown in Table\ref{tab:train_prompt}. Supervised fine-tune~(SFT) method is adopted to train these new tokens along with the LLM parameters, enabling them to adapt to the parameter distribution of the LLM. 


Formally, each instance consists of a prompt $C_{ui}$ along with a label $A_{ui}$. The objective of LLM is to modeling the following likelihood:

\begin{equation}
    \mathrm{LLM}(C_{ui}; \bar{\Theta}_{llm}) = P(A_{ui}|w_1, ..., w_{l_c}),
\end{equation}
where $\bar{\Theta}_{llm}$ represents the parameters of the LLM, including the user embedding matrix $U_e^L$, the item embedding matrix $I_e^L$, and other model-specific parameters, denoted as $\Theta_{llm}$.

Specifically, we employed two different answer types, each corresponding to a specific task in this paper. One answer type assumes that the tokens in label $A_{ui}$ belongs to natural language vocabulary $\mathcal{V}$, whereas the second answer type assumes that the tokens in label $A_{ui}$ belongs to the mixed vocabulary $\mathcal{W}$. Obviously, the latter task necessitates the model to generate the newly added user and item tokens, which is more demanding, but it has the potential to enhance user experience, for instance in conversational recommendation scenarios. 

The inputs are passed through the LLM to obtain the $l_c$th vector of $h_n$ denoted as $\mathcal{E}_{C_{ui}}$, which is then fed into the output layer to predict $A_{ui}$. And the loss function can be represented as follows:





\begin{equation}
    L_{llm} = -\log P(A_{ui}|w_1, ... , w_{l_c}).
\end{equation}


\subsection{Common Knowledge Enhanced Domain-Specific Recommendation Model}
Common embedding-based domain-specific models for personalized recommendation mainly rely on interactions between users and items to model user preferences and item characteristics. However, in many recommendation scenarios, the interaction data between users and items is often sparse, which requires common knowledge and reasoning abilities for better performance. 

Therefore, we tackle this problem using common knowledge through a joint approach with LLM. Specifically, we feed the prompt $C_{ui}$, which consists of interaction and contextual information into the LLM described in 4.1, and obtain it's top-layer feature $\mathcal{E}_{C_{ui}}$. This feature encompasses the LLM's knowledge and comprehension of user preferences and item characteristics.

Formally, the embedding based domain-specific recommendation~(DRS) models with consideration of the LLM knowledge can be formatted as

\begin{equation}
    \mathrm{DRS}(u, i; \bar{\Theta}_{drs}),
\end{equation}
where  $\bar{\Theta}_{drs}$ represents the parameters of the domain-specific RS model, including the user embedding matrix $U_e^D$, the item embedding matrix $I_e^D$, and other model-specific parameters, denoted as $\Theta_{drs}$.


We simply combine the LLM feature $\mathcal{E}_{C_{ui}}$ with the domain feature  $\mathcal{E}_{ui}^D$ extracted by domain-specific models through concatenation. As the training task of domain-specific model is to predict interaction rate, the output layer format and loss function can be represented as follows:



\begin{equation}
\begin{split}
    &\hat{y}_{\mathrm{drs}} = \sigma(\mathrm{MLP}(\mathtt{concat}(\mathcal{E}_{C_{ui}}, \mathcal{E}_{ui}^D))) \\
    &L_{drs} = 
    -[y \cdot \mathrm{log}(\hat{y}_{\mathrm{drs}}) 
    + (1-y) \cdot \mathrm{log}(1 - \hat{y}_{\mathrm{drs}})],
\end{split}
\end{equation}
where $\sigma$ represents the sigmoid function, $y$ equals $\mathcal{R}_{ui}$ and $L_{drs}$ represents the cross-entropy loss for the predicted $\hat{y}_{\mathrm{drs}}$ and the label $y$.

\subsection{Joint Model Learning}



\begin{algorithm}[h]
\caption{Algorithm of the proposed BDLM method} 
\label{alg:bdlm}  
\textbf{Input:} user set $\mathcal{U}$, item set $\mathcal{I}$, interactions set $\mathcal{R}$, prompt $C$, answer $A$, trade-off parameter $\gamma$, initial learning rate $\eta_1$ and $\eta_2$.

\textbf{Output:} LLM parameters $\bar{\Theta}_{llm} = \{\Theta_{llm}, U_e^L, I_e^L\}$ and domain-specific model parameters $\bar{\Theta}_{drs} = \{\Theta_{drs}, U_e^D, I_e^D\}$.

\textbf{Step1:} Train a domain RS model and get its user/item embeddings $DRS_U^0$ and $DRS_I^0$.

\textbf{Step2:} Extend user/item tokens for LLM and initialize all parameters $\bar{\Theta}_{llm}$, $\bar{\Theta}_{drs}$ and $\Theta_M = \{M_U, M_I\}$.

\Indp
$\Theta_{llm}$ is initialized with pretrained LLM parameters.

$U_e^L \gets DRS_U^0$; $I_e^L \gets DRS_I^0$;

$\Theta_{drs}$, $U_e^D$ and $I_e^D$ are initialized randomly.

$M_U \gets DRS_U^0$; $M_I \gets DRS_I^0$;

\Indm

\textbf{Step3:} Joint learning.

\Indp

\While{$\mathrm{stop\ condition\ is\ not\ reached}$}{

    Sample $(u, i)\in \mathcal{R}$, prompt $C_{ui}$, label $y$ and supervised answer $A_{ui}$. 
    
    \# \textbf{Training the LLM side.}
    
    Calculate $L_{llm}$ and $L_{m1}$ following Equation (2)-(4) and (7), then $\bar{L}_{llm} \gets L_{llm} + \gamma L_{m1}$;

    Update $M_u \gets u_e^L$, $M_i \gets i_e^L$;

    Update $\bar{\Theta}_{llm} \gets \bar{\Theta}_{llm} - \eta_1 \cdot \frac{\partial{\bar{L}_{llm}}}{\bar{\Theta}_{llm}}$ 

    \# \textbf{Training the DRS side.}

    Calculate $\mathcal{E}_{C_{ui}}$;

    Calculate $L_{drs}$ and $L_{m_2}$ following Equation (5)-(6) and (8) then $\bar{L}_{drs} \gets L_{drs} + \gamma L_{m2}$;

    Update $M_u \gets u_e^D$, $M_i \gets i_e^D$;

    Update $\bar{\Theta}_{drs} \gets \bar{\Theta}_{drs} - \eta_2 \cdot \frac{\partial{\bar{L}_{drs}}}{\bar{\Theta}_{drs}}$ 
    
  }
\end{algorithm}

The previous sections 4.1 and 4.2 introduced the method for providing static information supplementation. However, to further align the LLM with the domain-specific model and form a dynamic information sharing loop, we adopt a deep mutual learning approach to jointly train the LLM and the domain-specific model.

To be specific, as LLM and domain-specific model can be fomulated as $\mathrm{LLM}(C_{ui}; \bar{\Theta}_{llm})$ and $\mathrm{DRS}(u, i; \bar{\Theta}_{drs})$, one training sample in LLM and domain-specific model can be aligned in a unified form $\mathbf{x} = \{u, i, C_{ui}; y, A_{ui}\}$ for joint learning. 

To facilitate information sharing between LLM and domain-specific models through their common structure, i.e. user/item embedding layer, we introduce an information sharing module $M$.  During training, the LLM and the domain-specific model take turns updating a portion of the parameters $M_U$ and $M_I$ in the information sharing module. Mutual learning losses $L_{m1}$ and $L_{m2}$ are introduced during the updates to constrain the embedding differences between the LLM and the domain-specific model, as shown in the equations bellow.

\begin{equation}
    L_{m1} = || u_e^L - M_{u} ||_2 + || i_e^L - M_{i} ||_2
\end{equation}

\begin{equation}
    L_{m2} = || u_e^D - M_{u} ||_2 + || i_e^D - M_{i} ||_2,
\end{equation}
where $u_e^L$ and $i_e^L$ represent the embeddings of user $u$ and item $i$ in LLM, respectively. Similarly, $u_e^D$, $i_e^D$, $M_u$ and $M_i$ represent the corresponding embeddings in domain-specific model and information sharing module M. In addition, $||\cdot||_2$ represents MSE loss function.
 
By doing so, the embeddings of LLMs and domain-specific models are aligned in the same space, which can achieves effective information sharing. The total loss of joint learning can be represented as:

\begin{equation}
    L = L_{llm} + L_{drs} + \gamma(L_{m1} + L_{m2}),
\end{equation}
where $\gamma$ is a trade-off parameter acting as the weight of the mutual learning loss.

The complete training procedure is illustrated by the pseudocode as Algorithm~\ref{alg:bdlm}.

\section{Experiments}

In this section, we aim to evaluate the performance and effectiveness of the proposed method. Specifically, we conduct several experiments to study the following research questions:

\begin{itemize}
\item[$\bullet$] \textbf{RQ1:} Whether the proposed method BDLM can outperform SOTA recommendation models, including domain-specific models and LLM-based models.
\item[$\bullet$] \textbf{RQ2:} Whether the proposed method BDLM method sufficiently generalizable to enable information complementarity with different kinds of embedding-based domain-specific models?
\item[$\bullet$] \textbf{RQ3:} Whether the proposed method BDLM benefits from extending user/item tokens, preloading and deep mutual learning strategies?
\item[$\bullet$] \textbf{RQ4:} How do different configurations of the key hyper-parameters impact the performance of BDLM?
\end{itemize}

\subsection{Experimental Setup}
\subsubsection{Datasets}

We evaluated the proposed method BDLM on three widely used RS datasets from different fields, including "Movielens-1M", "Amazon Grocery and Gourmet Food" and "Amazon Health and Personal Care". We remove users and items with a low number of interactions as \cite{he2017neural}. The statistics of datasets are shown in Table~\ref{tab:data-stat}.

\begin{table}[t]\small
\caption{Statistics of experimental datasets.}
\centering
\begin{tabular}{lcccc}

\toprule
Dataset & \#Users & \#Items & \#Interactions & Sparsity \\
\midrule
MovieLens-1M & 6,040 & 3,952 & 1,000,224 & 0.958\\
Amazon-Grocery & 3,472 & 7,171 & 76,592 & 0.997\\
Amazon-Health & 4,872 & 7,934 & 107,135 & 0.997\\

\bottomrule

\end{tabular}
\label{tab:data-stat}
\end{table}


\begin{table}[t]\small
\caption{Prompt format of the interaction prediction task and the top-K recommendation task.}
\centering
\small

\begin{tabular}{p{0.18\linewidth}p{0.48\linewidth}c}
\toprule
Task & Instruction & Answer\\
\midrule
Interaction Prediction & User \red{<uid1>} has watched the following movies \red{<iid745>} (A Close Shave), \red{<iid588>} (Aladdin)... Predict if he/she will watch \red{<iid48>} (Pocahontas) next. Answer: & Yes\\
\midrule
Top-K Recommendation & User \red{<uid1>} has watched the following movies \red{<iid745>} (A Close Shave), \red{<iid588>} (Aladdin)... Predict which 1 movie in candidate set will he/she watch most probably? Candidates: \red{<iid1184>} (Mediterraneo), \red{<iid48>} (Pocahontas), \red{<iid1119>} (Drunks)... Answer: & \red{<iid48>} \\
\bottomrule
\end{tabular}
\label{tab:train_prompt}
\end{table}

\subsubsection{Metrics}
We evaluate the model's ability to capture user preferences through two tasks, namely \textbf{interaction prediction} and \textbf{top-K recommendation}. The prompt format of these two tasks is demonstrated as Table~\ref{tab:train_prompt}. We report precision, recall and f1 score for the interaction prediction task denoted as \textbf{prec}, \textbf{rec} and \textbf{f1} respectively. We report the top-1 and the top-2 recommendation hit rate for top-K recommendation task denoted as \textbf{HR@1} and \textbf{HR@2}. 

\begin{table*}[t]\small
\caption{Performance comparison of different methods. } 
\centering
\tabcolsep=0.16cm
\begin{tabular}{lccccccccccccccc}
\toprule
 & \multicolumn{5}{c}{ML-1M} & \multicolumn{5}{c}{Grocery} & \multicolumn{5}{c}{Health} \\
Models & HR@1 & HR@2 & prec & rec & f1 & HR@1 & HR@2 & prec & rec & f1 & HR@1 & HR@2 & prec & rec & f1 \\
\midrule
NCF              & 0.251 & 0.416 & 0.753 & 0.644 & 0.694 & 0.265 & 0.439 & 0.758 & 0.561 & 0.645 & 0.251 & 0.399 & 0.721 & 0.529 & 0.610 \\
GMF             & 0.265 & 0.416 & 0.756 & 0.685 & 0.719 & 0.326 & 0.482 & 0.744 & 0.570 & 0.646 & 0.316 & 0.435 & 0.722 & 0.525 & 0.608 \\
SASRec          & 0.324 & 0.475 & 0.748 & 0.644 & 0.691 & 0.317 & 0.465 & 0.753 & 0.569 & 0.648 & 0.325 & 0.442 & 0.740 & 0.551 & 0.632 \\
GRU4Rec         & 0.348 & 0.514 & 0.746 & 0.660 & 0.701 & 0.299 & 0.445 & 0.724 & 0.580 & 0.644 & 0.319 & 0.470 & 0.737 & 0.527 & 0.614 \\
BERT4Rec        & \underline{0.375} & \underline{0.506} & 0.740 & 0.668 & 0.702 & 0.256 & 0.403 & 0.721 & 0.536 & 0.615 & 0.265 & 0.396 & 0.701 & 0.534 & 0.606\\
LGCN             & 0.284 & 0.448 & 0.733 & \underline{0.702} & 0.717 & 0.337 & \underline{0.516} & \underline{0.764} & \underline{0.599} & \underline{0.672} & \underline{0.344} & \underline{0.481} & 0.739 & \underline{0.595} & \underline{0.659} \\

\midrule
LLM-Rec       & 0.233 & 0.388 & 0.649 & 0.580 & 0.612 & 0.079 & 0.011 & 0.545 & 0.246 & 0.339 & 0.092 & 0.119 & 0.605 & 0.236 & 0.339   \\
GPT4SM        & 0.245 & 0.408 & 0.666 & 0.664 & 0.665 & 0.233 & 0.388 & 0.649 & 0.580 & 0.612 & 0.244 & 0.378 & 0.633 & 0.534 & 0.580 \\
InstructRec      & 0.372 & 0.483 & \underline{0.833} & 0.672 & \underline{0.744} & \underline{0.339} & 0.437 & 0.761 & 0.567 & 0.650 & 0.309 & 0.358 & \underline{\textbf{0.770}} & 0.503 & 0.608 \\
\midrule
BDLM-llm   & 0.431 & 0.545 & \textbf{0.847} & 0.758 & 0.799 & 0.373 & 0.458 & 0.764 & \textbf{0.656} & \textbf{0.706} & 0.380 & 0.466 & 0.746 & 0.604 & 0.678 \\
BDLM-drs    & \textbf{0.460} & \textbf{0.607} & 0.812 & \textbf{0.819} & \textbf{0.816} & \textbf{0.440} & \textbf{0.567} & \textbf{0.780} & 0.632 & 0.698 & \textbf{0.403} & \textbf{0.546} & \textbf{0.770} & \textbf{0.626} & \textbf{0.690} \\
(Improv.) &  (22.7\%) & (20.0\%) & (1.7\%) & (16.7\%) & (9.7\%) & (29.8\%) & (9.9\%) & (1.7\%) & (9.5\%) & (5.1\%) & (17.2\%) & (13.5\%) & (0.0\%) & (5.2\%) & (4.7\%) \\
\bottomrule

\end{tabular}

\label{tab:main-res}
\end{table*}

\subsection{Implementation}
\subsubsection{Implementation of LLM}
We conduct our LLM experiments using Vicuna-7B \cite{zheng2023judging}, an SFT version of LLaMa-7B \cite{touvron2023llama}, which has 32 transformer layers, an embedding size of 4,096 dimensions, and a vocabulary size of 32,000 tokens. We use this SFT version instead of the native LLaMa because it has exhibited a certain level of contextual understanding and reasoning ability. This allows for faster convergence during fine-tuning for recommendation tasks.

We use Adam as the optimizer of LLM, and set the initial learning rate as 1e-5 with cosine annealing scheduler. We use 2 Nvidia A800 GPUs to perform full-parameter SFT on the LLM. We adopt the zero-3 strategy, where each GPU has a batch size of 8, resulting in a total batch size of 16 without gradient accumulation.


We expand the vocabulary based on the number of users and items in different datasets. For the three datasets, the vocabulary sizes have been expanded from the original 32,000 to 41,994, 42,645, and 44,808 respectively.

\subsubsection{Implementation of Domain-specific Models}

Domain-specific models are employed using two classical algorithms of RS, i.e. NCF \cite{he2017neural} and lightGCN \cite{he2020lightgcn}. These models are used to verify whether the community behavior pattern information is helpful for improving the recommendation performance of the LLM. In order to align with the configuration of the LLM for preloading and collaborative training, we set the user and item embedding size of the domain-specific models to 4096. 

We use Adam as the optimizer for the training of domain-specific models, and the initial learning rate is set to 1e-4. The batch-size is also set to 16 for collaborative training with LLM, and the trade-off parameter $\gamma$ of mutual learning loss is set within the range [0, 1e-3, 1e-2, 1e-1, 1, 10].

\subsubsection{Evaluation Settings}

Both tasks use leave-one-out strategy for partitioning training and testing sets, but the construction of the testing set differs. For the interaction prediction task, which involves binary classification, we sample an equal number of negative samples with the positive ones. For the top-K recommendation task, the candidate items are not the entire set of products, due to the input text length limitation of LLM. Thus, referring to the methods InstructRec~\citep{DBLP:journals/corr/abs-2305-07001}, a sample of items that users have not interacted with are selected as negative samples. The test set consists of 20 candidate items, including 1 positive instance and 19 negative instances. Additionally, in order to increase the difficulty of negative instances in the testing set, a combination of popularity sampling and text similarity sampling was used during negative sampling.

\subsection{Baselines}
We utilize two groups of methods as our baselines which involve state-of-the-art domain-specific models for recommendation system and LLM-based methods for recommendation system.

(1) Domain-specific models:
\textbf{NCF} \cite{he2017neural}: A neural network based collaborative filtering method, which can extract community behavior pattern information through massive interaction data.

\textbf{GMF} \cite{he2017neural}: Generalized matrix factorization, which adopts element-wise multiplication between user and item embeddings and then generate the prediction through MLP.

\textbf{LGCN} \cite{he2020lightgcn}: LightGCN, a GCN-based state-of-the-art method for extracting interaction graph information for users and items.

\textbf{GRU4Rec} \cite{hidasi2015session}: An RNN-based model for session-based recommendations, which extract user behavior pattern by recurrent neural network.

\textbf{SASRec} \cite{kang2018self}: A state-of-the-art method for sequential recommendation based on self-attention layers.

\textbf{BERT4Rec} \cite{sun2019bert4rec}: A state-of-the-art method based on a bi-directional self-attentive model with the cloze objective for sequential recommendation.

(2) LLM-based methods:

\textbf{LLM-Rec}~\citep{lyu2023llm}: A non-SFT and prompt-based method for personalized recommendation. By designing appropriate prompts, they stimulated the emergence capacity of the LLM, resulting in improved performance. We implement this method using ChatGPT.

\textbf{InstructRec}~\citep{DBLP:journals/corr/abs-2305-07001}: A SFT-based method, thoroughly analyzed various tasks and information in the field of recommendation systems and comprehensively designed various instructions.

\textbf{GPT4SM} \cite{peng2023gpt}: A domain-specific model is enhanced by incorporating knowledge information from LLM. The top-layer embedding from LLM is fed into the domain-specific model.




\subsection{Main Results}
Table~\ref{tab:main-res} shows the performance of different methods on the three benchmark datasets. The numbers in bold denote the best results among all methods and the underlined numbers denote the best results among baselines. BDLM-drs denotes the results obtained from domain-specific model side whereas BDLM-llm represents the results from LLM side. We choose lightGCN as the domain-specific model because it archives the best experimental results among all domain-specific models. From the experimental results, we can get the following conclusions for RQ1.

\begin{itemize}
    \item Firstly, our proposed BDLM method outperforms the baseline models on all datasets, which demonstrates the effectiveness of the proposed method. Specifically, our method improve the best baseline by 22.7\%, 29.8\% and 17.2\% on the three datasets measured by \textbf{HR@1}.
    \item Secondly, it is noteworthy that the text-only LLMs methods (InstructRec) that do not extend the user and item tokens, demonstrate a relatively good performance when dealing with movie recommendation tasks, but perform poorly on e-commence scenarios. This is because the pre-trained LLM has more external knowledge about movies and can accurately differentiate between various movies based on their names.  However, in an e-commerce scenario, the LLM cannot accurately distinguish the features of products based solely on similar textual descriptions. As a result, it cannot provide precise recommendations to users.
    \item Moreover, LLM-enhanced domain-specific model method (GPT4SM) does not present a good performance. This result demonstrates that contextual information extracted by LLM without instruction fine-tuning may not provide useful information for domain-specific models, and could potentially introduce noise. 
    \item Finally, in the two tasks of the experiments, BDLM achieves a higher improvement ratio in the top-K recommendation task compared to the interaction prediction task. This is because the former requires extracting the user's preference from a large number of candidate items with similar descriptions, which is more challenging than the latter task. BDLM fully utilizes the information complementarity strategy to achieve a higher improvement over the baseline on even more difficult problems.
\end{itemize}

\subsection{Analysis}

\subsubsection{Generalization Study (RQ2)}

In order to verify the sufficient generalizability of the proposed BDLM method, which enables information complementarity with different kinds of embedding-based domain-specific models, we also conducted experiments with the classical NCF algorithm, where all instances of lightGCN in the BDLM approach are replaced with NCF model. 

The experimental results are shown in Table 4. Our proposed BDLM-NCF approach demonstrates superior performance compared to both NCF and text-only LLM approach ~(InstructRec) across all datasets, which proves the generalizability of BDLM.

\begin{table}[t]\small
\caption{Experimental results of BDLM-NCF on MovieLens-1M, Amazon-Grocery and Amazon-Health test sets for generalization study.}
\centering
\tabcolsep=0.16cm
\begin{tabular}{lcccccc}
\toprule
 & \multicolumn{2}{c}{ML-1M} & \multicolumn{2}{c}{Grocery} & \multicolumn{2}{c}{Health} \\
Models & HR@1 & f1 & HR@1 & f1 & HR@1 & f1 \\
\midrule
NCF              & 0.251 & 0.694 & 0.265 & 0.645 & 0.251 & \underline{0.610} \\
InstructRec      & \underline{0.372} & \underline{0.744} & \underline{0.339} & \underline{0.650} & \underline{0.309} & 0.608 \\
\midrule
BDLM-NCF-llm     & 0.398 & 0.789 & 0.387 & 0.677 & 0.331 & 0.648  \\
BDLM-NCF-drs         & \textbf{0.405} & \textbf{0.800} & \textbf{0.396} & \textbf{0.683} & \textbf{0.361} & \textbf{0.664} \\
(Improv.) &  (8.9\%) & (7.5\%) & (16.8\%) & (5.1\%) & (16.8\%) & (8.9\%) \\
\bottomrule

\end{tabular}

\label{tab:generalization}
\end{table}

\subsubsection{Ablation Study (RQ3)}

\begin{figure}[h]
    \centering
    \includegraphics[width=8.8cm]{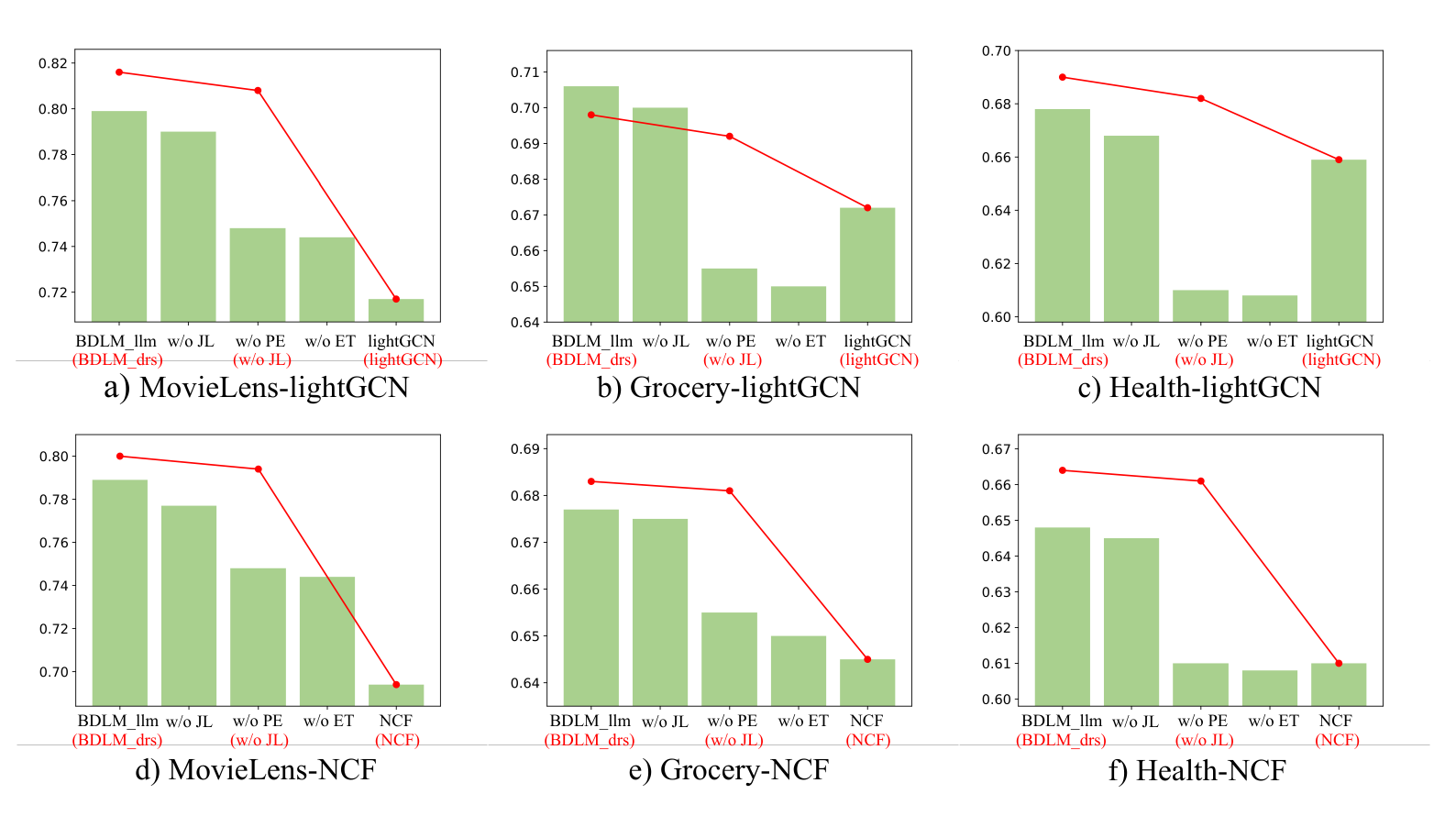}
    \caption{The ablation study on different datasets with lightGCN and NCF as domain-specific model. The green bar represents the ablations for the LLM, while the red line illustrates those of the domain-specific model.}
    \label{fig:ablation}
\end{figure}

We conduct comprehensive ablation experiments for better understanding the contributions of different components in our BDLM approach. The ablation experimental groups are designed as follows.

\textbf{BDLM\_LGCN\_drs} \& \textbf{BDLM\_NCF\_drs}: The final results from domain-specific model side based on lightGCN and NCF, respectively.

\textbf{BDLM\_LGCN\_llm} \& \textbf{BDLM\_NCF\_llm}: The final LLM-side results based on LGCN and NCF, respectively.

\textbf{w/o JL}: The results without \textbf{J}oint \textbf{L}earning.

\textbf{w/o PE}: The results without \textbf{P}re-loading user and item \textbf{E}mbeddings.

\textbf{w/o ET}: The results without \textbf{E}xending user and item \textbf{T}okens.

The results of ablation study are shown in Figure~\ref{fig:ablation}. From the experimental results, we can get the following conclusions.

\begin{itemize}
    \item[-] Each component of the proposed method contributes to the overall improvement in the recommendation task. Experimental results indicate that if the extended tokens are randomly initialized (w/o PE), LLM will face difficulty in converging, and thus its performance is nearly equivalent to not extending the tokens (w/o ET). However, when the domain-specific model is introduced and the extended tokens of LLM are initialized with the pre-trained embeddings, the performance of LLM shows a significant improvement (from w/o PE to w/o JL). This suggests that domain information extracted by the domain-specific model can be transmitted to LLM through user and item embeddings.
    \item[-] Through joint learning, LLM can better utilize the community behavior pattern information provided by the domain-specific model, and the knowledge and reasoning information generated by LLM has stronger adaptability to the domain-specific model. As a result, it further enhances the final performance (from w/o JL to BDLM).
\end{itemize}

\subsubsection{Hyper-parameter analysis (RQ4)}

\begin{figure}[t]
    \centering
    \includegraphics[width=8.5cm]{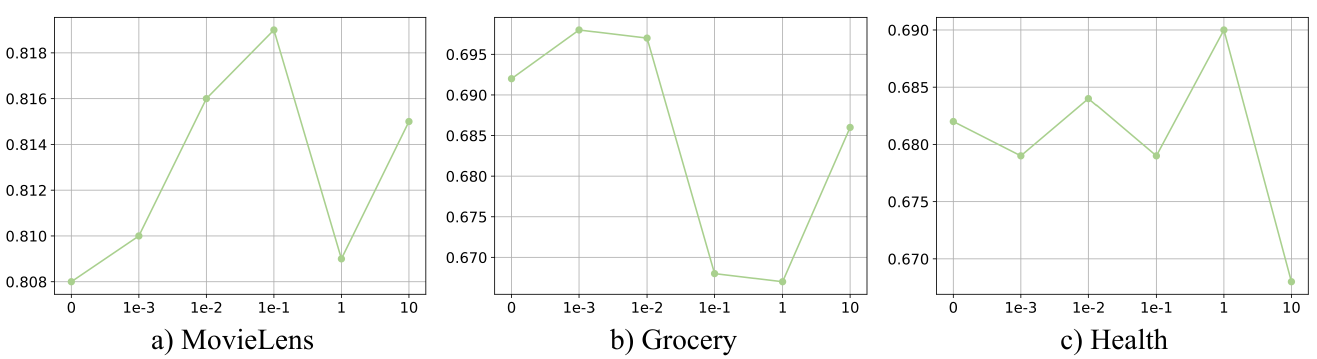}
    \caption{Effects of trade-off parameter $\gamma$.}
    \label{fig:hyper}
\end{figure}

This section analyzes the impact of the trade-off parameter $\gamma$ in the loss function on the effectiveness of joint training. We conducted experiments on three datasets with $\gamma$ values ranging from [0, 1e-3, 1e-2, 1e-1, 1, 10]. The experimental results are shown in the Figure~\ref{fig:hyper}. It can be observed that across the three datasets, as the $\gamma$ increases, the experimental performance generally exhibits an initial improvement followed by a decline. The choice of $\gamma$ varies for different datasets. The best $\gamma$ for three datasets are 1e-1, 1e-3 and 1 respectively.


\section{Conclusion}

In this paper, we propose a unified paradigm for bridging the information gap between domain-specific models and generative large language models~(BDLM) for personalized recommendation. Domain-specific models can improve the performance of LLMs by providing community behavior pattern information, while LLMs can contribute their general knowledge and reasoning capabilities to domain-specific models. Results of comparison experiments with state-of-the-art baseline methods demonstrate the effectiveness of the proposed information complementarity strategy. The ablation study demonstrates the significance of each component of BDLM.

In this paper, we only explored the recommendation performance in multi-shot scenarios. In future work, we will continue to leverage the knowledge and reasoning capabilities of LLM, as well as the alignment of LLMs and domain-specific models, to optimize recommendations in new user and new item scenarios.

\bibliographystyle{ACM-Reference-Format}
\bibliography{sample-base}


\end{document}